\documentclass[11pt, a4paper]{article}

\usepackage[margin=1in]{geometry}
\usepackage{tabularx}
\newcolumntype{Y}{>{\centering\arraybackslash}X}
\usepackage{graphicx}
\usepackage{booktabs}
\usepackage{multirow}
\usepackage[table]{xcolor}
\usepackage{xcolor}
\usepackage{float}
\usepackage{amsmath,amssymb,amsfonts,amsthm}
\usepackage{mathrsfs}
\usepackage{textcomp}
\usepackage[version=4]{mhchem}
\usepackage[T1]{fontenc}
\usepackage[utf8]{inputenc}
\usepackage{authblk}
\usepackage{fontawesome5}
\usepackage[numbers,sort&compress]{natbib}

\usepackage[colorlinks=true,linkcolor=blue,citecolor=blue,urlcolor=blue]{hyperref}

\usepackage{algorithm}
\usepackage{algorithmicx}
\usepackage{algpseudocode}
\usepackage{listings}

\theoremstyle{plain}

\theoremstyle{definition}

\begin{document}

\title{MLIP-MC: A Framework for Adsorption Simulations using Machine-Learned Interatomic Potentials}

\author[1]{Connor W. Edwards}
\author[1]{Fengxu Yang}
\author[1]{Konstantin Stracke}
\author[1]{Jack D. Evans\thanks{Corresponding author: j.evans@adelaide.edu.au}}

\affil[1]{School of Physics, Chemistry and Earth Sciences, Adelaide University, North Terrace, Adelaide, 5005, South Australia, Australia}

\date{}

\maketitle

\begin{abstract}
Grand canonical Monte Carlo (GCMC) simulations are essential for screening metal–organic frameworks (MOFs) for gas adsorption, yet their accuracy is limited by underlying interatomic potentials. Universal machine-learned interatomic potentials (MLIPs), trained on diverse chemical datasets, promise zero-shot prediction without system-specific training. We introduce MLIP-MC, an open-source Python framework to conduct GCMC simulations with MLIPs, and use this framework to benchmark a series of universal models, including MACE-MP-0, ORB-v3, and fairchem ODAC, for \ce{CO2} adsorption on ZIF-8, ZIF-4, and Mg-MOF-74. All universal models exhibit systematic biases, consistently over- or underestimating adsorption energetics. Crucially, accuracy depends on training data composition: only models trained on MOF-adsorbate interactions achieve reasonable agreement with a density functional theory derived reference. Errors grow linearly with \ce{CO2} uptake, reflecting compounding inaccuracies in adsorbate–adsorbate interactions. Our results demonstrate that current universal MLIPs require finetuning for quantitative adsorption predictions and demonstrate the power of MLIP-MC to rapidly test models.
\end{abstract}

\newpage

\section{Introduction}
Gas adsorption in porous materials underpins a diverse range of industrial processes and emerging technologies critical to modern society.
Since the commercialisation of synthetic zeolites in the 1950s, adsorption-based separations have become essential to petroleum refining, air separation, and chemical processing.\cite{10.1021/acs.chemrev.7b00738}
Metal-organic frameworks (MOFs) have emerged as a revolutionary class of porous adsorbents.\cite{Furukawa2013}
Constructed from metal clusters and organic linkers, MOFs offer unprecedented tunability: in principle, an almost unlimited number of distinct frameworks can be synthesized by varying the building blocks.
This structural diversity creates exciting opportunities for optimising materials for specific applications, from carbon capture and natural gas storage to drug delivery and catalysis.\cite{Mahajan2022, Nath2024, AbnadesLzaro2020, LinderPatton2025}

High-throughput computational screening is a powerful research tool to address the challenge of optimising MOFs.
Snurr and coworkers demonstrated that molecular simulations can rapidly evaluate the adsorption properties of thousands of MOF structures, enabling the discovery of structure-property relationships and identification of top-performing candidates prior to synthesis.\cite{10.1039/c4cs00070f}
The construction of computation-ready databases such as CoRE-MOF, containing over $10\,000$ experimentally characterized structures, has accelerated these efforts.\cite{10.1021/cm502594j}
Large-scale screening of hypothetical MOFs has revealed key structure-property relationships for methane storage and \ce{CO2} capture.\cite{10.1038/nchem.1192}
Grand Canonical Monte Carlo (GCMC) simulations, combined with classical force fields such as UFF (Universal Force Field) or DREIDING, now routinely screen databases of hypothetical and experimental MOFs.\cite{10.1080/08927022.2015.1010082}
However, the accuracy of these screening studies is fundamentally limited by the underlying force fields or classical potentials.
Generic potentials like UFF were parametrised for organic molecules and often fail to capture the subtle electronic effects that govern adsorption on MOFs, particularly the strong interactions at open metal sites or the cooperative binding observed in flexible frameworks.\cite{10.1021/ja00051a040}
This limitation has motivated extensive efforts to develop more accurate descriptions of host-guest interactions through ab initio methods.

Sauer and coworkers have pioneered the application of high-level quantum chemical methods to achieve high accuracy (errors below 4~kJ$\,$mol$^{-1}$) for adsorption on zeolites and MOFs.\cite{10.1021/acs.accounts.9b00506}
For example, their hybrid MP2 and density functional theory (DFT) approach combines periodic DFT for the extended framework with correlated wavefunction methods (MP2, coupled cluster) for the local adsorption site, enabling quantitative prediction of adsorption enthalpies for small molecules including water, alkanes, and \ce{CO2}.\cite{10.1021/acs.jpcc.5b01739}
These calculations have provided benchmark-quality data for understanding water-zeolite interactions,\cite{10.1021/acs.jpcc.1c04270} alcohol adsorption, and catalytic reaction mechanisms, while also revealing the limitations of standard DFT functionals for describing dispersion-dominated physisorption.\cite{10.1021/acs.accounts.9b00506}

Despite these advances, ab initio methods remain computationally prohibitive for the extensive configurational sampling required in Monte Carlo simulations.
A single DFT energy evaluation for a MOF unit cell may require minutes to hours, and GCMC simulations demand millions of energy calculations.
This computational gap has limited the impact of ab initio accuracy on practical screening studies.
Machine-learned interatomic potentials (MLIPs) offer a compelling solution to bridge this gap, achieving near-DFT accuracy at a fraction of the computational cost.\cite{arxiv.2601.16459, Edwards2025}
Goeminne et al. demonstrated that MLIPs trained on high-level DFT interaction energies can enable accurate GCMC simulations of MOFs, in comparison to experiment, deriving adsorption isotherms for \ce{CO2} on both ZIF-8 and the open-metal site containing Mg-MOF-74.\cite{10.1021/acs.jctc.3c00495}
However, this approach requires system-specific training data for each MOF of interest.

Recent developments have produced several universal MLIPs trained on massive datasets spanning diverse chemical space.\cite{10.1038/s41524-024-01500-6}
These models offer the potential for zero-shot prediction, without system-specific retraining.
These include MACE-MP-0, trained on the Materials Project trajectory dataset using higher-order equivariant message passing\cite{10.1063/5.0297006}; the Orb family of models, employing graph network simulators optimised for inference speed\cite{10.48550/ARXIV.2410.22570, 10.48550/ARXIV.2504.06231}; and the fairchem UMA models, leveraging mixture-of-linear-experts architectures to unify multiple application domains including catalysis, bulk materials, and, critically, direct air capture through the ODAC datasets.\cite{10.5281/zenodo.15587498, 10.1021/acscentsci.3c01629}

In this work, we introduce MLIP-MC, an open-source Python framework that enables Monte Carlo simulations of gas adsorption using machine-learned interatomic potentials.
We use this framework to systematically benchmark the leading universal MLIPs (orb-v3-conservative-inf-omat, MACE-MP-0a, MACE-MP-0b3, MACE-MPA-0, fairchem ODAC23, fairchem ODAC25) to predict the \ce{CO2} adsorption properties on three representative MOFs: ZIF-8, Mg-MOF-74, and ZIF-4.\cite{10.1073/pnas.0602439103, 10.1021/ja809459e}
These systems were chosen to span a range of adsorption mechanisms: ZIF-8 features purely physisorptive interactions in cage-like cavities, Mg-MOF-74 exhibits strong chemisorptive binding at open metal sites, and ZIF-4 represents a dense, narrow-pore framework.
We compare universal MLIP predictions against the reference finetuned model reported by Goeminne et al.,\cite{10.1021/acs.jctc.3c00495} trained on ab initio adsorption data, to quantify the accuracy and limitations of zero-shot prediction, and to assess whether universal MLIPs have reached sufficient maturity for practical adsorption screening applications.

\section{Computational methods}

\subsection{MLIP-MC framework}
MLIP-MC is implemented in Python and builds upon the Atomic Simulation Environment (ASE) for structure manipulation and calculator interfaces.\cite{10.1088/1361-648X/aa680e} 
The framework currently provides two primary simulation capabilities: Widom insertion calculations for determining isosteric heats of adsorption at infinite dilution \cite{10.1063/1.1734110} and GCMC simulations for computing adsorption isotherms.
Users can interact with the framework either through a robust command-line interface (\verb|mlip_mc|) for direct shell execution or via high-level Python wrappers such as \verb|run_gcmc()| and \verb|run_widom()| for programmatic control and seamless integration into larger computational workflows.
The GCMC implementation follows the standard Metropolis algorithm with molecule insertion, deletion, rotation, and translation moves.\cite{FrenkelSmit2002}
A rejection criterion is implemented should a move cause van der Waals overlap between the adsorbate and the framework, or other adsorbates. 
Van der Waals radii are taken from the default ASE values and, as reported by Goeminne et al., the predicted interaction energies appear insensitive to the chosen radii.\cite{10.1021/acs.jctc.3c00495}
This van der Waals overlap check is also applied during the attempted Widom insertion steps.

Thermodynamic consistency is maintained through the Peng-Robinson equation of state (PREOS), which computes the chemical potential and fugacity coefficients based on temperature and pressure conditions. A built-in table of critical parameters for common adsorbates is automatically applied during simulation.
The MLIP-MC framework supports both equilibration phases (where configurations are sampled but not recorded) and production phases (where ensemble averages are computed). 

During execution, MLIP-MC also constantly records restart data at every simulation step to ensure seamless recovery against wall-time limitations and unforeseen system interruptions.
Currently, MLIP-MC supports three MLIP backends: FAIRChem for models trained within the Open Catalyst framework, MACE-Torch for MACE-architecture models and Orbital for the Orbital models.\cite{fairchem, Batatia2022mace, Batatia2022Design, 10.48550/ARXIV.2504.06231} The framework was further adapted to support Nequip models, and a branch was implemented to run the finetuned models reported by Goeminne et al.\cite{10.1021/acs.jctc.3c00495}
The model weights can be loaded from local files or automatically downloaded from Hugging Face repositories.\cite{huggingface}
Pore size calculations were performed using ZEO++.\cite{zeo_paper}

\subsection{Simulation details}
Our MLIP-MC framework was used to perform Widom insertion and GCMC simulations of \ce{CO2} on ZIF-8, Mg-MOF-74, and ZIF-4. Overall, six universal models were benchmarked against data generated using a finetuned model provided by Goeminne et al.; MACE-MP-0a,\cite{10.1063/5.0297006} MACE-MP-0b3,\cite{10.1063/5.0297006} MACE-MPA-0,\cite{10.1063/5.0297006} ORB-v3-inf-consv-omat,\cite{10.48550/ARXIV.2504.06231} fairchem ODAC23,\cite{fairchem} and fairchem ODAC25\cite{10.48550/arXiv.2508.03162} (Table~\ref{tab:model_specs}).

\begin{table*}[htbp!]
    \centering
    \caption{Summary of universal MLIP models employed. Models are classified by architecture: Graph Neural Network (GNN), Graph Network Simulator (GNS), or Graph Transformer (GT). Training datasets and computational efficiency are also listed. Efficiency was computed on a single NVIDIA A100 GPU (SXM4, 40 GB).}
    \label{tab:model_specs}
    \resizebox{\textwidth}{!}{%
    \begin{tabular}{l |c |c |c |l}
        Name & DOI & Class & Efficiency (s/step) & Datasets\\
        \hline
        \hline
        ORB-v3-inf-consv-omat & 10.48550/ARXIV.2504.06231 & GNS & 0.0167 & OMat\cite{10.48550/arXiv.2410.12771}\\
        MACE-MP-0a & 10.1063/5.0297006 & GNN & 0.0081 & MPtrj\cite{mptrj_2023}\\
        MACE-MP-0b3 & 10.1063/5.0297006 & GNN & 0.0231 & MPtrj\cite{mptrj_2023} \\
        MACE-MPA-0 & 10.1063/5.0297006 & GNN & 0.0279 & MPtrj\cite{mptrj_2023} + sAlex\cite{salex_2024}\\
        {fairchem ODAC23} & 10.5281/zenodo.15587498 & GT & 0.0556 & ODAC23\cite{10.1021/acscentsci.3c01629}\\
        {fairchem ODAC25} (full) & 10.5281/zenodo.15587498 & GT & 0.1431 & ODAC25\cite{10.48550/arXiv.2508.03162}\\
    \end{tabular}
    }
\end{table*}

For the Widom insertion simulations, $100\,000$ insertion steps were performed with all insertion energies recorded for configurations in which the inserted \ce{CO2} molecule did not exhibit van der Waals overlap with the framework. Each simulation then provided the average Widom insertion energy, the Boltzmann-weighted average of the insertion energy and the isosteric heat of adsorption at zero coverage. The Widom insertion energy was used to compute an isosteric heat at infinite dilution using Equation~\ref{qst}:

\begin{equation}
Q_{st, dilution}
= - \frac{\left\langle \Delta E \exp\left(-\beta \Delta E\right) \right\rangle}
       {\left\langle \exp\left(-\beta \Delta E\right) \right\rangle}
+ RT ,
\label{qst}
\end{equation}

where $\beta = (k_{\mathrm{B}} T)^{-1}$, $k_{\mathrm{B}}$ is the Boltzmann constant, $T$ is the temperature, and $\Delta E$ denotes the \ce{CO2} insertion energy.

GCMC simulations were performed at 273~K (for ZIF-4 and ZIF-8) and 298~K (for Mg-MOF-74) for 14 pressures; 0.0002, 0.001, 0.005, 0.02, 0.05, 0.1, 0.2, 0.5, 1, 2, 5, 10, 20 and 50~bar.
Each simulation comprised $100\,000$ equilibration steps followed by $1\,000\,000$ production steps with an equal probability of a rotation, translation, insertion or deletion move. 
A van der Waals overlap check was employed to ensure the move did not cause overlap between adsorbate and framework atoms.
The accepted production steps were subsequently used to calculate the isosteric heat of adsorption using Equation~\ref{iso_heat}:

\begin{equation}
Q_{st}
= -\frac{
\left\langle E N \right\rangle
- \left\langle E \right\rangle \left\langle N \right\rangle
}{
\left\langle N^{2} \right\rangle
- \left\langle N \right\rangle^{2}
} + RT
\label{iso_heat}
\end{equation}
where $N$ is the number of adsorbed \ce{CO2} molecules and $E$ is the total interaction energy of the system. In this description, the isosteric heat captures both framework–adsorbate and adsorbate–adsorbate interactions.

\section{Results and discussion}
To test the MLIP-MC framework, six universal models were explored with diverse classes and training datasets (Table~\ref{tab:model_specs}). 
The training datasets vary in their relevance to adsorption.
MPtrj, OMat and sAlex contain DFT calculations of inorganic materials derived from the Materials Project and the Alexandria databases but include no adsorption data. The Open Direct Air Capture datasets (ODAC${23}$ and ODAC${25}$) were specifically constructed to examine gas adsorption, covering single adsorption and co-adsorption of \ce{CO2} and \ce{H2O} on MOFs. The ODAC${25}$ dataset extends the ODAC${23}$ to include structures with multiple adsorbate molecules, sampling higher \ce{CO2} uptake. However, the exact distribution of adsorbed \ce{CO2} molecules in the ODAC${25}$ is unclear.

The reference data for GCMC simulations (uptake and isosteric heat data) were taken from results reported by Goeminne et al. using their finetuned models (model\_ZIF8\_N=1000.pth, model\_MgMOF74\_pbed3.pth, model\_ZIF3\_4\_6\_8.pth).\cite{10.1021/acs.jctc.3c00495} As Widom insertion results were not provided, the MLIP-MC framework was adapted to support Nequip models (molmod branch) and used to create reference isosteric heats for each MOF using these finetuned models.

\subsection{Isosteric heat of adsorption at infinite dilution}

\begin{figure}[htbp!]
    \vspace{5mm}
    \centering
    \includegraphics[]{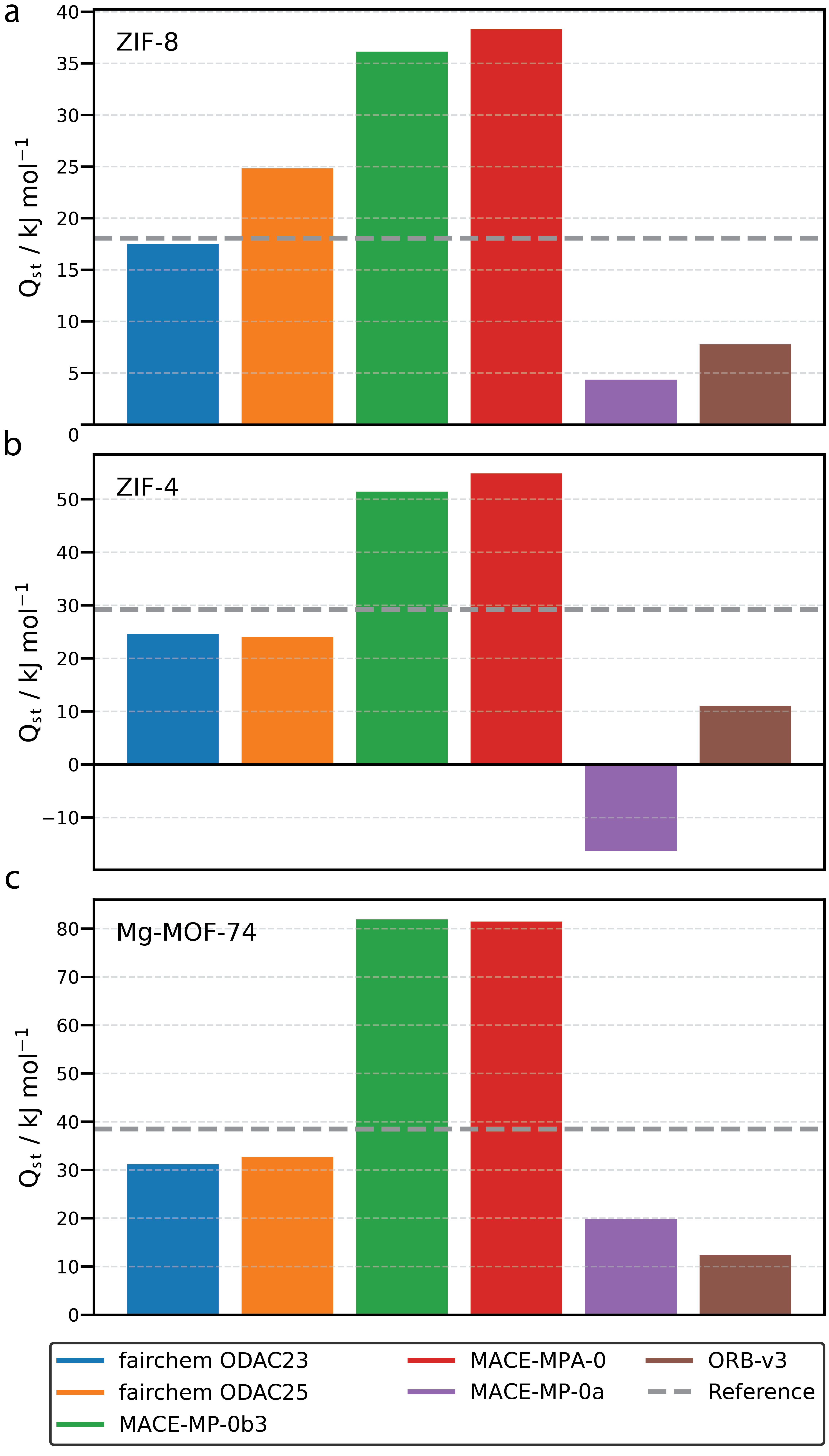}
    \caption{Isosteric heat of adsorption of \ce{CO2} at infinite dilution ($Q_{st, dilution}$) on a) ZIF-8, b) ZIF-4 and c) Mg-MOF-74 calculated by Widom insertion for a series of universal models. Reference data from the finetuned model of Goeminne et al.\cite{10.1021/acs.jctc.3c00495}}
    \label{fig:Widom}
    \vspace{5mm}
\end{figure}

The systematic trends in the isosteric heat at infinite dilution across the universal models are illustrated in Figure~\ref{fig:Widom}.
Rather than exhibiting framework-specific errors, the models display a systematic bias across all three MOFs, either over- or underestimating the reference isosteric heat values.

Unsurprisingly, the fairchem ODAC family of models shows the strongest overall agreement with the reference data. Both models generally underestimate the isosteric heat across all frameworks; however, the ODAC25 model slightly overestimates the predicted adsorption strength of ZIF-8, which has a large pore size and no open metal sites. In contrast, the MACE-based models exhibit large deviations from the reference. MACE-MP-0b3 and MACE-MPA-0 systematically overestimate the isosteric heat for all MOFs, with the magnitude increasing for Mg-MOF-74, containing open metal sites. This behaviour suggests a tendency of these models to exaggerate the MOF-\ce{CO2} interaction strength. Conversely, the MACE-MP-0a model consistently underestimates the isosteric heat of adsorption, even producing negative values for ZIF-4. Given that the MACE models share similar architectures, these results underscore the significance of training data for model accuracy. Similarly, ORB-v3 consistently underestimates the isosteric heat across all frameworks. 

Overall, no model achieves quantitative agreement with the reference Widom insertion results across all frameworks, though fairchem ODAC model family shows good agreement. The consistency of the observed over- and underbinding trends across the three MOFs indicates that these errors are governed by the nature and coverage of the training data rather than being a framework-dependent problem. As the ODAC/ODAC25 datasets include MOF-\ce{CO2} interactions, predictably models trained on these data generate the best results. However, these Widom results only involve one \ce{CO2} molecule; it is therefore important to consider how these results scale with the amount of \ce{CO2} adsorbed.

\subsection{Adsorption isotherms}
Models can be generally classified into two categories: those that systematically underestimate the interaction energies (fairchem ODAC, MACE-MP-0a and ORB-v3) and those that overestimate them (MACE-MPA-0, MACE-MP-0b3). The impact of these over- and underpredictions of the isosteric heat is directly reflected in the uptake of \ce{CO2}. Models that underestimate the isosteric heat also underestimate the \ce{CO2} uptake, as the reduced heat released upon adsorption makes the process less thermodynamically favourable. In contrast, models that overestimate the isosteric heat show higher \ce{CO2} uptakes, reflecting an artificially increased driving force for adsorption. 

\begin{figure}[htbp!]
    \vspace{5mm}
    \centering
    \includegraphics[]{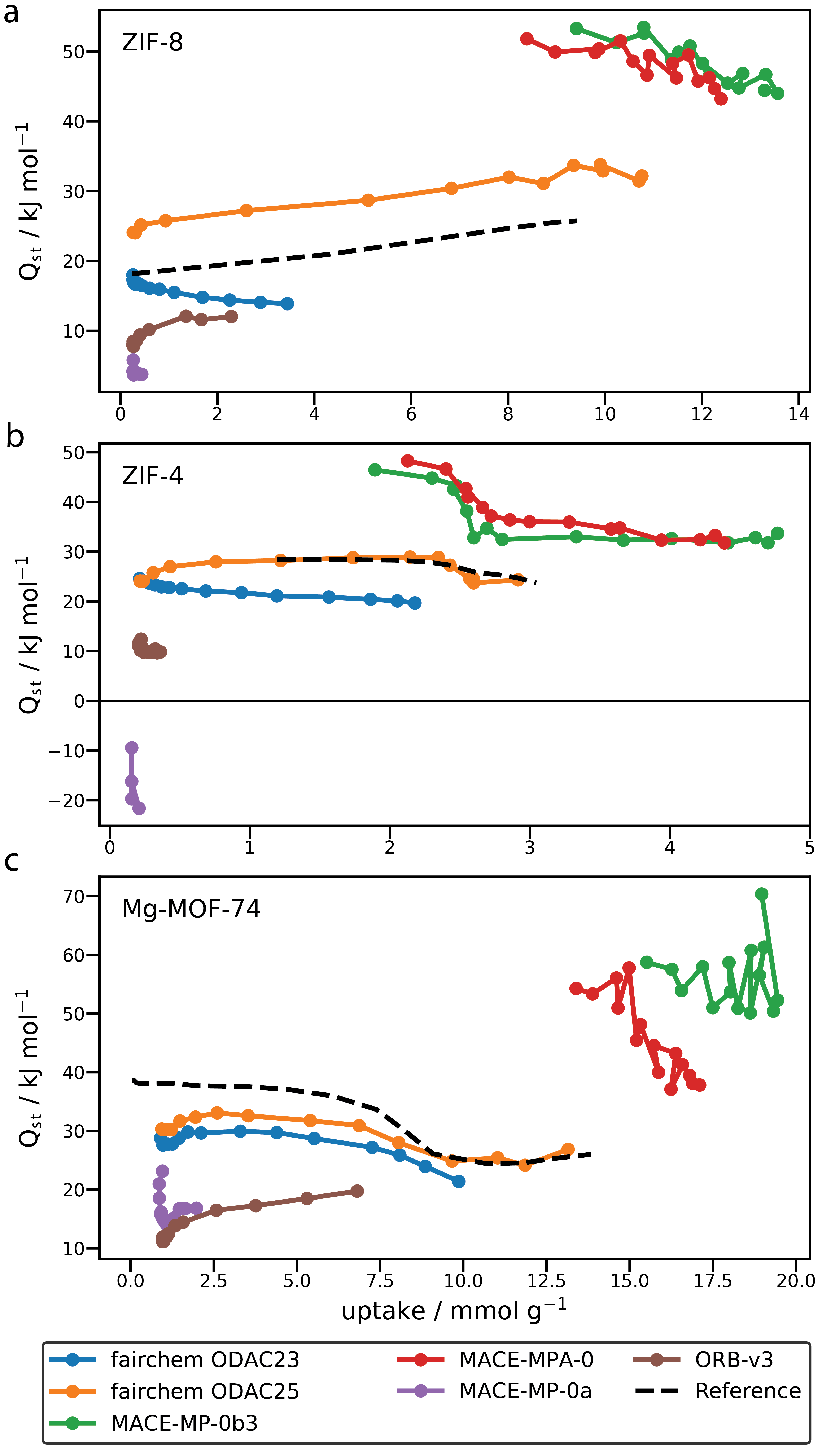}
    \caption{Isosteric heat of adsorption ($Q_{st}$) for \ce{CO2} on a) ZIF-8, b) ZIF-4 and c) Mg-MOF-74 computed by GCMC simulations for a series of universal models. Reference data from a finetuned model of Goeminne et al.\cite{10.1021/acs.jctc.3c00495}}
    \label{fig:isosteric_heat}
    \vspace{5mm}
\end{figure}

The MACE models exhibit extreme systematic biases in the predicted isosteric heat across all investigated MOFs and pressures (Figure~\ref{fig:isosteric_heat}). Because MACE-MPA-0 and MACE-MP-0b3 significantly overestimate the isosteric heat, adsorption becomes too favourable. As a consequence, these models completely filled the pores for all MOFs, even at low pressures. Conversely, MACE-MP-0a had extremely low isosteric heats, even giving negative values for ZIF-4. This means it is thermodynamically unfavourable to adsorb \ce{CO2}, causing the pores to be empty. Increasing pressure did not significantly impact the \ce{CO2} uptake for this model, indicating that MACE-MP-0a is unable to capture the pressure dependence of adsorption. ORB-v3 similarly underestimates the isosteric heat across all pressures, although the bias is less severe than that seen for MACE-MP-0a. Notably, none of these models are able to reproduce the distinctive pressure-dependent shape of the isosteric heat of adsorption for Mg-MOF-74, related to the filling of open metal sites.

\begin{figure}[htbp!]
    \vspace{5mm}
    \centering
    \includegraphics[]{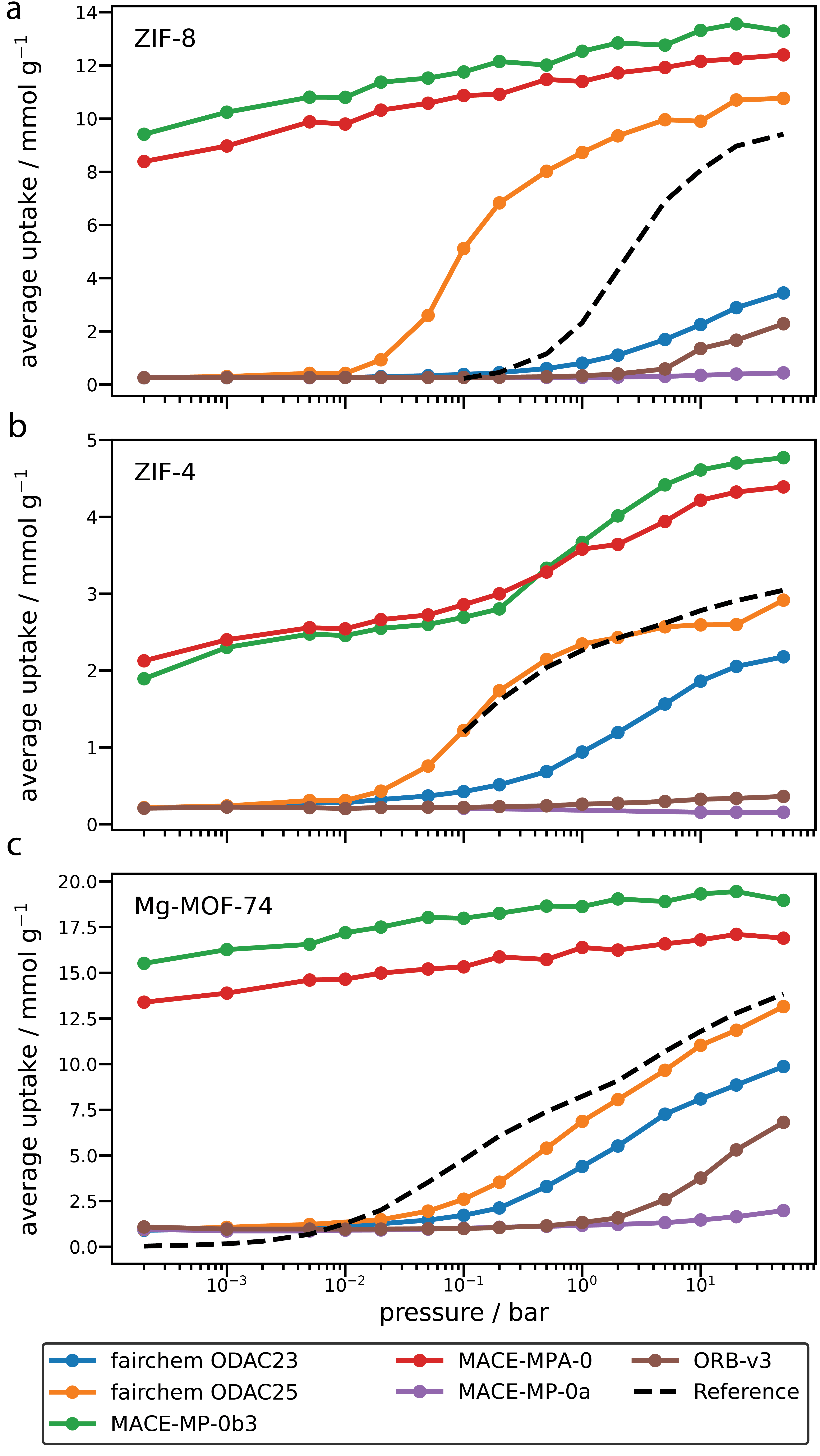}
    \caption{\ce{CO2} adsorption isotherms for a) ZIF-8, b) ZIF-4 and c) Mg-MOF-74 calculated using GCMC simulations for a series of universal models. Reference data from the finetuned model of Goeminne et al.\cite{10.1021/acs.jctc.3c00495}}
    \label{fig:uptake}
    \vspace{5mm}
\end{figure}

Consistent with the Widom results, the fairchem ODAC models show a clear improvement over the MACE and ORB models, being able to approximately simulate the adsorption isotherm (Figure~\ref{fig:uptake}) for ZIF-4 and Mg-MOF-74 without either leaving the pore empty or completely filling the pore. While fairchem ODAC had reasonable predictions for ZIF-8 at lower pressures, it fails to capture the correct adsorption behaviour at higher pressures. This behaviour likely arises from the limitations of the ODAC23 dataset, which contains only structures with up to one adsorbed \ce{CO2} molecule. As a result, the model underestimates \ce{CO2}–\ce{CO2} interactions and is not able to treat higher uptakes. Of the models evaluated, only the fairchem ODAC25 model is able to reproduce all key features of the reference data for ZIF-4 and Mg-MOF-74. However, fairchem ODAC25 fails to quantitatively describe ZIF-8, systematically overestimating the isosteric heat of adsorption.

This discrepancy likely arises from an inaccurate representation of framework–\ce{CO2} interactions. Although ZIF-8 and Mg-MOF-74 have comparable pore sizes, 11.8~\AA and 11.7~\AA\ respectively (computed by ZEO++), their adsorption mechanisms differ substantially. Mg-MOF-74 contains open metal sites that strongly bind \ce{CO2}. This leads to a high isosteric heat at low pressures, which decreases as these sites become saturated. In contrast, ZIF-8 lacks open metal sites and adsorption is dominated by weaker dispersive interactions. Consequently, ZIF-8 is expected to exhibit lower isosteric heat than Mg-MOF-74, as reflected in the reference data and fairchem ODAC25 results. 
However, fairchem ODAC25 systematically overestimates the isosteric heat of adsorption for ZIF-8 across all pressures, suggesting the model overemphasises pore size as a driving factor for adsorption. For Mg-MOF-74, the isosteric heat is underestimated at lower uptakes, suggesting the binding strength to open metal sites is underestimated. As pressure and \ce{CO2} uptake increase, these open metal sites become saturated and \ce{CO2} starts filling the pore (Supplementary Figure~1). As this happens, the model more closely reproduces the reference behaviour. While this apparent agreement at higher uptakes is encouraging, it likely arises from partial error cancellation between underestimated metal–\ce{CO2} interactions and overestimated \ce{CO2}-\ce{CO2} interactions within the large pore. This behaviour highlights that fairchem ODAC25 does not yet robustly capture the distinct physical adsorption mechanisms governing these two frameworks.

Overall, no single model provides universally accurate adsorption energetics across all regimes. The ability to recreate the reference data is governed primarily by the training data describing \ce{CO2}-framework interactions rather than by architectural complexity. Training on datasets with multiple adsorbates such as ODAC25 showed improvements over datasets with just a single adsorbate such as ODAC23. This enables models to better describe \ce{CO2}-\ce{CO2} interactions and give a more reliable description of the adsorption behaviour. From the models evaluated here, fairchem ODAC25 demonstrates the most reliable and transferable behaviour for predicting both adsorption energetics and \ce{CO2} uptake, yet falls short for ZIF-8.

\subsection{Shortcomings of universal models}
\begin{figure}[htbp!]
    \vspace{5mm}
    \centering
    \includegraphics[]{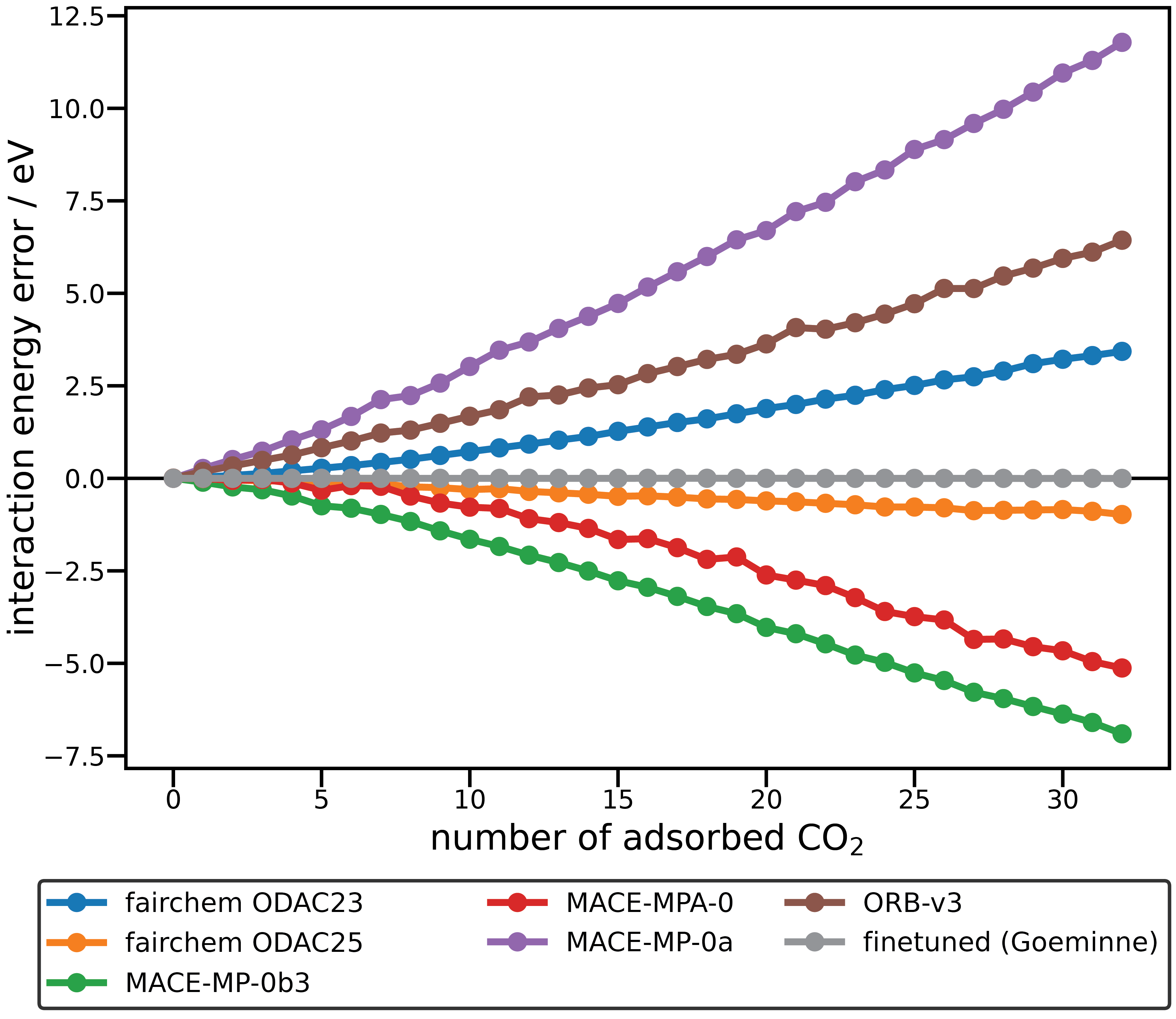}
    \caption{Error in interaction energy, model predicted relative to the reference DFT reported by Goeminne et al.\cite{10.1021/acs.jctc.3c00495}, with increasing \ce{CO2} molecules for the adsorption on ZIF-8.}
    \label{fig:error_w_co2}
    \vspace{5mm}
\end{figure}

The work so far establishes that select universal models can capture realistic isosteric heats at infinite dilution; however, only fairchem ODAC25 can recreate the work conducted by Goeminne et al.\cite{10.1021/acs.jctc.3c00495} for ZIF-4 and Mg-MOF-74, but not ZIF-8.
To further investigate where this discrepancy arises, the ZIF-8 DFT reference data reported by Goeminne et al. (employed to train the ZIF-8 finetuned model) was used to validate the universal models and the finetuned model (Figure~\ref{fig:error_w_co2}). This dataset consists of $30\,000$ ZIF-8 structures with \ce{CO2} uptakes ranging from 0 to 32 molecules and between 551 and 3663 structures per uptake. 

As expected, the model finetuned on this dataset reproduced the DFT interaction energies across all uptakes. In contrast, the universal models all exhibited uptake-dependent error. While their performance is reasonable at low uptake, the error in interaction energy grows approximately linearly with increasing \ce{CO2} uptake. Consistent with the Widom and GCMC simulations, MACE-MP-0a, ORB-v3 and fairchem ODAC overestimate the interaction energies and MACE-MPA-0, MACE-MP-0b3 underestimate the insertion energies. Fairchem ODAC25 performs significantly better than other universal models across all uptakes. However, fairchem ODAC25 still slightly underestimates the interaction energy.

The near linear growth in interaction error suggests the accuracy of these models has a strong dependence on the \ce{CO2} uptake. At low uptakes, the level of error is dependent on the ability of the model to describe framework-\ce{CO2} interactions. As the uptake increases, \ce{CO2}–\ce{CO2} interactions become more prevalent and have a larger contribution to the total error. The linear trend indicates that both interaction regimes are described with comparable inaccuracy. There is a subtle increase in error around 8~\ce{CO2} molecules (most noticeable in MACE-MPA-0) which corresponds to when \ce{CO2}-\ce{CO2} interactions begin to dominate. This suggests that overall \ce{CO2}-\ce{CO2} interactions are represented less accurately than framework-\ce{CO2} interactions.

\ce{CO2} interaction error becomes more evident when considering the error per \ce{CO2}, which retains a linear increase for most models (Supplementary Figure~2). As the number of \ce{CO2} molecules grows, so too does the number of \ce{CO2}–\ce{CO2} interactions, leading to compounding errors. Notably, fairchem ODAC25 exhibits a flat error per \ce{CO2} profile, suggesting a systematic offset in \ce{CO2}–\ce{CO2} interactions rather than the loading-dependent errors seen in the other models. This improved behaviour arises from its training dataset, which includes structures with multiple adsorbates, enabling a more accurate representation of intermolecular interactions at higher uptakes. However, even fairchem ODAC25 fails to quantitatively reproduce reference interaction energies, highlighting the need for system-specific finetuned models.

Beyond accuracy, the fastest universal model evaluated (MACE-MP-0a) demonstrates comparable computational efficiency to the finetuned Mg-MOF-74 model. In contrast, both fairchem models are significantly less efficient than the other universal models, with the ODAC25 model being 14 times slower than the finetuned model. As a result, even though fairchem ODAC25 was the only universal model to reproduce isotherm behaviours, the high computational cost limits its practicality for longer simulations or extensive high-pressure screening studies. This further highlights the importance of developing finetuned models that can achieve both accuracy and computational efficiency. 

The efficiency metrics reported in Table~\ref{tab:model_specs} should be treated with caution, as efficiency will depend on the amount of adsorbed \ce{CO2}. Models that predict little to no uptake, such as MACE-MP-0a, will appear artificially more efficient than models that predict higher uptakes. Because the number of atoms varies during the simulation, these efficiencies have not been normalised on a per-atom basis. Instead, they reflect the real-world production efficiency of applying each model to Mg-MOF-74 at 1~bar and 273~K.

\section{Summary and Outlook}
In this work, we introduced MLIP-MC, an open-source Python framework that enables Widom insertions and GCMC simulations using universal MLIPs. We benchmarked six universal models against a reference dataset obtained from a series of finetuned models trained for \ce{CO2} adsorption on ZIF-8, ZIF-4 and Mg-MOF-74.

Across Widom and GCMC simulations, all MLIPs exhibited systematic biases in their interaction energies. Rather than being system-specific, these biases resulted from the models under- or overestimating the \ce{CO2} interaction energies. Consequently, MACE-MPA-0 and MACE-MP-0b3 artificially filled the available pores, while MACE-MP-0a, ORB-v3 and fairchem ODAC produced partially or fully empty pores across all pressures. Of the models evaluated, only fairchem ODAC25 showed reliable results, successfully reproducing the reference isotherms for ZIF-4 and Mg-MOF-74. However, fairchem ODAC25 was unable to properly describe \ce{CO2} uptake in ZIF-8 due to overemphasising the role of \ce{CO2}-\ce{CO2} interactions, and thus pore size, in the absence of open metal sites. Additionally, the computational inefficiency of the fairchem ODAC models limits their practical applicability.

These results demonstrate that existing universal MLIPs are unable to reliably describe gas adsorption across diverse systems and that finetuned models are required to capture the correct thermodynamic behaviour of adsorbed gases. Incorporating more adsorption structures in training datasets and extending these to cover systems beyond one adsorbed \ce{CO2}, such as ODAC25, significantly improves the accuracy of these models. This highlights the importance of high-quality and diverse training data for future MLIPs. The MLIP-MC framework enables benchmarking and the deployment and training of new MLIPs. We expect the next generation of accurate and efficient models will enable high-throughput discovery and screening of new porous materials using MLIPs for gas separation and storage applications.

\section*{Supporting Information}
MLIP-MC is available on GitHub \href{https://github.com/jackevansadl/MLIP-MC}{\faGithub}. All code and data for this research are available on Zenodo: \href{https://doi.org/10.5281/zenodo.18637262}{10.5281/zenodo.18637262}. Figures of the \ce{CO2} distribution on Mg-MOF-74 and error analysis are provided in the Supplementary Information.

\section*{Acknowledgements}
J.D.E. is the recipient of an Australian Research Council Discovery Early Career Award (project number DE220100163) funded by the Australian Government. Phoenix HPC service at Adelaide University is thanked for providing high-performance computing resources. This research was supported by the Australian Government’s National Collaborative Research Infrastructure Strategy (NCRIS), with access to computational resources provided by Pawsey Supercomputing Research Centre through the National Computational Merit Allocation Scheme.


\begin{thebibliography}{37}
\providecommand{\natexlab}[1]{#1}
\providecommand{\url}[1]{\texttt{#1}}
\expandafter\ifx\csname urlstyle\endcsname\relax
  \providecommand{\doi}[1]{doi: #1}\else
  \providecommand{\doi}{doi: \begingroup \urlstyle{rm}\Url}\fi

\bibitem[Dusselier and Davis(2018)]{10.1021/acs.chemrev.7b00738}
Michiel Dusselier and Mark~E. Davis.
\newblock Small-pore zeolites: Synthesis and catalysis.
\newblock \emph{Chemical Reviews}, 118\penalty0 (11):\penalty0 5265–5329, May
  2018.
\newblock ISSN 1520-6890.
\newblock \doi{10.1021/acs.chemrev.7b00738}.
\newblock URL \url{http://dx.doi.org/10.1021/acs.chemrev.7b00738}.

\bibitem[Furukawa et~al.(2013)Furukawa, Cordova, O’Keeffe, and
  Yaghi]{Furukawa2013}
Hiroyasu Furukawa, Kyle~E. Cordova, Michael O’Keeffe, and Omar~M. Yaghi.
\newblock The chemistry and applications of metal-organic frameworks.
\newblock \emph{Science}, 341\penalty0 (6149), August 2013.
\newblock ISSN 1095-9203.
\newblock \doi{10.1126/science.1230444}.
\newblock URL \url{http://dx.doi.org/10.1126/science.1230444}.

\bibitem[Mahajan and Lahtinen(2022)]{Mahajan2022}
Shreya Mahajan and Manu Lahtinen.
\newblock Recent progress in metal-organic frameworks (mofs) for co2 capture at
  different pressures.
\newblock \emph{Journal of Environmental Chemical Engineering}, 10\penalty0
  (6):\penalty0 108930, December 2022.
\newblock ISSN 2213-3437.
\newblock \doi{10.1016/j.jece.2022.108930}.
\newblock URL \url{http://dx.doi.org/10.1016/j.jece.2022.108930}.

\bibitem[Nath et~al.(2024)Nath, Wright, Ahmed, Siegel, and Matzger]{Nath2024}
Karabi Nath, Keenan~R. Wright, Alauddin Ahmed, Donald~J. Siegel, and Adam~J.
  Matzger.
\newblock Adsorption of natural gas in metal–organic frameworks: Selectivity,
  cyclability, and comparison to methane adsorption.
\newblock \emph{Journal of the American Chemical Society}, 146\penalty0
  (15):\penalty0 10517–10523, April 2024.
\newblock ISSN 1520-5126.
\newblock \doi{10.1021/jacs.3c14535}.
\newblock URL \url{http://dx.doi.org/10.1021/jacs.3c14535}.

\bibitem[Abánades~Lázaro et~al.(2020)Abánades~Lázaro, Wells, and
  Forgan]{AbnadesLzaro2020}
Isabel Abánades~Lázaro, Connor J.~R. Wells, and Ross~S. Forgan.
\newblock Multivariate modulation of the zr mof uio‐66 for
  defect‐controlled combination anticancer drug delivery.
\newblock \emph{Angewandte Chemie International Edition}, 59\penalty0
  (13):\penalty0 5211–5217, February 2020.
\newblock ISSN 1521-3773.
\newblock \doi{10.1002/anie.201915848}.
\newblock URL \url{http://dx.doi.org/10.1002/anie.201915848}.

\bibitem[Linder-Patton et~al.(2025)Linder-Patton, Wang, Evans, Yasin,
  Berahim-Jusoh, Li, Huang, Phak, Seman, Sumby, and Doonan]{LinderPatton2025}
Oliver~M. Linder-Patton, Lizhuo Wang, Jack~D. Evans, Nor~Hafizah Yasin,
  Nor~Hafizah Berahim-Jusoh, Siqi Li, Jun Huang, Chan~Zhe Phak, Akbar~A. Seman,
  Christopher~J. Sumby, and Christian~J. Doonan.
\newblock Understanding the role of the zr-mof support structure on templated
  ternary co 2 hydrogenation catalyst structure and activity.
\newblock \emph{ACS Applied Materials \& Interfaces}, 17\penalty0
  (31):\penalty0 44573–44584, July 2025.
\newblock ISSN 1944-8252.
\newblock \doi{10.1021/acsami.5c10085}.
\newblock URL \url{http://dx.doi.org/10.1021/acsami.5c10085}.

\bibitem[Colón and Snurr(2014)]{10.1039/c4cs00070f}
Yamil~J. Colón and Randall~Q. Snurr.
\newblock High-throughput computational screening of metal–organic
  frameworks.
\newblock \emph{Chem. Soc. Rev.}, 43\penalty0 (16):\penalty0 5735–5749, 2014.
\newblock ISSN 1460-4744.
\newblock \doi{10.1039/c4cs00070f}.
\newblock URL \url{http://dx.doi.org/10.1039/c4cs00070f}.

\bibitem[Chung et~al.(2014)Chung, Camp, Haranczyk, Sikora, Bury,
  Krungleviciute, Yildirim, Farha, Sholl, and Snurr]{10.1021/cm502594j}
Yongchul~G. Chung, Jeffrey Camp, Maciej Haranczyk, Benjamin~J. Sikora, Wojciech
  Bury, Vaiva Krungleviciute, Taner Yildirim, Omar~K. Farha, David~S. Sholl,
  and Randall~Q. Snurr.
\newblock Computation-ready, experimental metal–organic frameworks: A tool to
  enable high-throughput screening of nanoporous crystals.
\newblock \emph{Chemistry of Materials}, 26\penalty0 (21):\penalty0
  6185–6192, October 2014.
\newblock ISSN 1520-5002.
\newblock \doi{10.1021/cm502594j}.
\newblock URL \url{http://dx.doi.org/10.1021/cm502594j}.

\bibitem[Wilmer et~al.(2011)Wilmer, Leaf, Lee, Farha, Hauser, Hupp, and
  Snurr]{10.1038/nchem.1192}
Christopher~E. Wilmer, Michael Leaf, Chang~Yeon Lee, Omar~K. Farha, Brad~G.
  Hauser, Joseph~T. Hupp, and Randall~Q. Snurr.
\newblock Large-scale screening of hypothetical metal–organic frameworks.
\newblock \emph{Nature Chemistry}, 4\penalty0 (2):\penalty0 83–89, November
  2011.
\newblock ISSN 1755-4349.
\newblock \doi{10.1038/nchem.1192}.
\newblock URL \url{http://dx.doi.org/10.1038/nchem.1192}.

\bibitem[Dubbeldam et~al.(2015)Dubbeldam, Calero, Ellis, and
  Snurr]{10.1080/08927022.2015.1010082}
David Dubbeldam, Sofía Calero, Donald~E. Ellis, and Randall~Q. Snurr.
\newblock Raspa: molecular simulation software for adsorption and diffusion in
  flexible nanoporous materials.
\newblock \emph{Molecular Simulation}, 42\penalty0 (2):\penalty0 81–101,
  February 2015.
\newblock ISSN 1029-0435.
\newblock \doi{10.1080/08927022.2015.1010082}.
\newblock URL \url{http://dx.doi.org/10.1080/08927022.2015.1010082}.

\bibitem[Rappe et~al.(1992)Rappe, Casewit, Colwell, Goddard, and
  Skiff]{10.1021/ja00051a040}
A.~K. Rappe, C.~J. Casewit, K.~S. Colwell, W.~A. Goddard, and W.~M. Skiff.
\newblock Uff, a full periodic table force field for molecular mechanics and
  molecular dynamics simulations.
\newblock \emph{Journal of the American Chemical Society}, 114\penalty0
  (25):\penalty0 10024–10035, December 1992.
\newblock ISSN 1520-5126.
\newblock \doi{10.1021/ja00051a040}.
\newblock URL \url{http://dx.doi.org/10.1021/ja00051a040}.

\bibitem[Sauer(2019)]{10.1021/acs.accounts.9b00506}
Joachim Sauer.
\newblock Ab initio calculations for molecule–surface interactions with
  chemical accuracy.
\newblock \emph{Accounts of Chemical Research}, 52\penalty0 (12):\penalty0
  3502–3510, November 2019.
\newblock ISSN 1520-4898.
\newblock \doi{10.1021/acs.accounts.9b00506}.
\newblock URL \url{http://dx.doi.org/10.1021/acs.accounts.9b00506}.

\bibitem[Piccini et~al.(2015)Piccini, Alessio, Sauer, Zhi, Liu, Kolvenbach,
  Jentys, and Lercher]{10.1021/acs.jpcc.5b01739}
GiovanniMaria Piccini, Maristella Alessio, Joachim Sauer, Yuchun Zhi, Yuanshuai
  Liu, Robin Kolvenbach, Andreas Jentys, and Johannes~A. Lercher.
\newblock Accurate adsorption thermodynamics of small alkanes in zeolites. ab
  initio theory and experiment for h-chabazite.
\newblock \emph{The Journal of Physical Chemistry C}, 119\penalty0
  (11):\penalty0 6128–6137, March 2015.
\newblock ISSN 1932-7455.
\newblock \doi{10.1021/acs.jpcc.5b01739}.
\newblock URL \url{http://dx.doi.org/10.1021/acs.jpcc.5b01739}.

\bibitem[Stanciakova et~al.(2021)Stanciakova, Louwen, Weckhuysen, Bulo, and
  G\"{o}ltl]{10.1021/acs.jpcc.1c04270}
Katarina Stanciakova, Jaap~N. Louwen, Bert~M. Weckhuysen, Rosa~E. Bulo, and
  Florian G\"{o}ltl.
\newblock Understanding water–zeolite interactions: On the accuracy of
  density functionals.
\newblock \emph{The Journal of Physical Chemistry C}, 125\penalty0
  (37):\penalty0 20261–20274, September 2021.
\newblock ISSN 1932-7455.
\newblock \doi{10.1021/acs.jpcc.1c04270}.
\newblock URL \url{http://dx.doi.org/10.1021/acs.jpcc.1c04270}.

\bibitem[Edwards et~al.(2026)Edwards, Linder-Patton, and
  Evans]{arxiv.2601.16459}
Connor~W. Edwards, Oliver~M. Linder-Patton, and Jack~D. Evans.
\newblock Simulations of high temperature decomposition of metal-organic
  frameworks to form amorphous catalysts, 2026.
\newblock URL \url{https://arxiv.org/abs/2601.16459}.

\bibitem[Edwards and Evans(2025)]{Edwards2025}
Connor~W. Edwards and Jack~D. Evans.
\newblock Exploring foundational machine learned potentials for treating the
  high temperature dynamics of metal‐organic frameworks.
\newblock \emph{Advanced Theory and Simulations}, 9\penalty0 (2), October 2025.
\newblock ISSN 2513-0390.
\newblock \doi{10.1002/adts.202500514}.
\newblock URL \url{http://dx.doi.org/10.1002/adts.202500514}.

\bibitem[Goeminne et~al.(2023)Goeminne, Vanduyfhuys, Van~Speybroeck, and
  Verstraelen]{10.1021/acs.jctc.3c00495}
Ruben Goeminne, Louis Vanduyfhuys, Veronique Van~Speybroeck, and Toon
  Verstraelen.
\newblock Dft-quality adsorption simulations in metal–organic frameworks
  enabled by machine learning potentials.
\newblock \emph{Journal of Chemical Theory and Computation}, 19\penalty0
  (18):\penalty0 6313–6325, August 2023.
\newblock ISSN 1549-9626.
\newblock \doi{10.1021/acs.jctc.3c00495}.
\newblock URL \url{http://dx.doi.org/10.1021/acs.jctc.3c00495}.

\bibitem[Deng et~al.(2025)Deng, Choi, Zhong, Riebesell, Anand, Li, Jun,
  Persson, and Ceder]{10.1038/s41524-024-01500-6}
Bowen Deng, Yunyeong Choi, Peichen Zhong, Janosh Riebesell, Shashwat Anand,
  Zhuohan Li, KyuJung Jun, Kristin~A. Persson, and Gerbrand Ceder.
\newblock Systematic softening in universal machine learning interatomic
  potentials.
\newblock \emph{npj Computational Materials}, 11\penalty0 (1), January 2025.
\newblock ISSN 2057-3960.
\newblock \doi{10.1038/s41524-024-01500-6}.
\newblock URL \url{http://dx.doi.org/10.1038/s41524-024-01500-6}.

\bibitem[Batatia et~al.(2025)Batatia, Benner, Chiang, Elena, Kovács,
  Riebesell, Advincula, Asta, Avaylon, Baldwin, Berger, Bernstein, Bhowmik,
  Bigi, Blau, Cărare, Ceriotti, Chong, Darby, De, Della~Pia, Deringer,
  Elijošius, El-Machachi, Fako, Falcioni, Ferrari, Gardner, Gawkowski,
  Genreith-Schriever, George, Goodall, Grandel, Grey, Grigorev, Han, Handley,
  Heenen, Hermansson, Ho, Hofmann, Holm, Jaafar, Jakob, Jung, Kapil, Kaplan,
  Karimitari, Kermode, Kourtis, Kroupa, Kullgren, Kuner, Kuryla, Liepuoniute,
  Lin, Margraf, Magdău, Michaelides, Moore, Naik, Niblett, Norwood, O’Neill,
  Ortner, Persson, Reuter, Rosen, Rosset, Schaaf, Schran, Shi, Sivonxay,
  Stenczel, Sutton, Svahn, Swinburne, Tilly, van~der Oord, Vargas,
  Varga-Umbrich, Vegge, Vondrák, Wang, Witt, Wolf, Zills, and
  Csányi]{10.1063/5.0297006}
Ilyes Batatia, Philipp Benner, Yuan Chiang, Alin~M. Elena, Dávid~P. Kovács,
  Janosh Riebesell, Xavier~R. Advincula, Mark Asta, Matthew Avaylon, William~J.
  Baldwin, Fabian Berger, Noam Bernstein, Arghya Bhowmik, Filippo Bigi,
  Samuel~M. Blau, Vlad Cărare, Michele Ceriotti, Sanggyu Chong, James~P.
  Darby, Sandip De, Flaviano Della~Pia, Volker~L. Deringer, Rokas Elijošius,
  Zakariya El-Machachi, Edvin Fako, Fabio Falcioni, Andrea~C. Ferrari, John
  L.~A. Gardner, Mikołaj~J. Gawkowski, Annalena Genreith-Schriever, Janine
  George, Rhys E.~A. Goodall, Jonas Grandel, Clare~P. Grey, Petr Grigorev,
  Shuang Han, Will Handley, Hendrik~H. Heenen, Kersti Hermansson, Cheuk~Hin Ho,
  Stephan Hofmann, Christian Holm, Jad Jaafar, Konstantin~S. Jakob, Hyunwook
  Jung, Venkat Kapil, Aaron~D. Kaplan, Nima Karimitari, James~R. Kermode,
  Panagiotis Kourtis, Namu Kroupa, Jolla Kullgren, Matthew~C. Kuner, Domantas
  Kuryla, Guoda Liepuoniute, Chen Lin, Johannes~T. Margraf, Ioan-Bogdan
  Magdău, Angelos Michaelides, J.~Harry Moore, Aakash~A. Naik, Samuel~P.
  Niblett, Sam~Walton Norwood, Niamh O’Neill, Christoph Ortner, Kristin~A.
  Persson, Karsten Reuter, Andrew~S. Rosen, Louise A.~M. Rosset, Lars~L.
  Schaaf, Christoph Schran, Benjamin~X. Shi, Eric Sivonxay, Tamás~K. Stenczel,
  Christopher Sutton, Viktor Svahn, Thomas~D. Swinburne, Jules Tilly, Cas
  van~der Oord, Santiago Vargas, Eszter Varga-Umbrich, Tejs Vegge, Martin
  Vondrák, Yangshuai Wang, William~C. Witt, Thomas Wolf, Fabian Zills, and
  Gábor Csányi.
\newblock A foundation model for atomistic materials chemistry.
\newblock \emph{The Journal of Chemical Physics}, 163\penalty0 (18), November
  2025.
\newblock ISSN 1089-7690.
\newblock \doi{10.1063/5.0297006}.
\newblock URL \url{http://dx.doi.org/10.1063/5.0297006}.

\bibitem[Neumann et~al.(2024)Neumann, Gin, Rhodes, Bennett, Li, Choubisa,
  Hussey, and Godwin]{10.48550/ARXIV.2410.22570}
Mark Neumann, James Gin, Benjamin Rhodes, Steven Bennett, Zhiyi Li, Hitarth
  Choubisa, Arthur Hussey, and Jonathan Godwin.
\newblock Orb: A fast, scalable neural network potential, 2024.
\newblock URL \url{https://arxiv.org/abs/2410.22570}.

\bibitem[Rhodes et~al.(2025)Rhodes, Vandenhaute, Šimkus, Gin, Godwin, Duignan,
  and Neumann]{10.48550/ARXIV.2504.06231}
Benjamin Rhodes, Sander Vandenhaute, Vaidotas Šimkus, James Gin, Jonathan
  Godwin, Tim Duignan, and Mark Neumann.
\newblock Orb-v3: atomistic simulation at scale, 2025.
\newblock URL \url{https://arxiv.org/abs/2504.06231}.

\bibitem[Shuaibi et~al.()Shuaibi, Das, Sriram, Misko, Barroso-Luque, Gao,
  Goyal, Ulissi, Wood, Xie, Yoon, Wander, Kolluru, Barnes, Sunshine, Tran,
  Xiang, Levine, Shoghi, Chair, , Lan, Tian, Musielewicz, clz55, Hu, , Michel,
  willis, and vbttchr]{10.5281/zenodo.15587498}
Muhammed Shuaibi, Abhishek Das, Anuroop Sriram, Misko, Luis Barroso-Luque, Ray
  Gao, Siddharth Goyal, Zachary Ulissi, Brandon Wood, Tian Xie, Junwoong Yoon,
  Brook Wander, Adeesh Kolluru, Richard Barnes, Ethan Sunshine, Kevin Tran,
  Xiang, Daniel Levine, Nima Shoghi, Ilias Chair, , Janice Lan, Kaylee Tian,
  Joseph Musielewicz, clz55, Weihua Hu, , Kyle Michel, willis, and vbttchr.
\newblock {FAIRChem}.
\newblock URL \url{https://github.com/facebookresearch/fairchem}.

\bibitem[Sriram et~al.(2024)Sriram, Choi, Yu, Brabson, Das, Ulissi,
  Uyttendaele, Medford, and Sholl]{10.1021/acscentsci.3c01629}
Anuroop Sriram, Sihoon Choi, Xiaohan Yu, Logan~M. Brabson, Abhishek Das,
  Zachary Ulissi, Matt Uyttendaele, Andrew~J. Medford, and David~S. Sholl.
\newblock The open dac 2023 dataset and challenges for sorbent discovery in
  direct air capture.
\newblock \emph{ACS Central Science}, 10\penalty0 (5):\penalty0 923–941, May
  2024.
\newblock ISSN 2374-7951.
\newblock \doi{10.1021/acscentsci.3c01629}.
\newblock URL \url{http://dx.doi.org/10.1021/acscentsci.3c01629}.

\bibitem[Park et~al.(2006)Park, Ni, C\^oté, Choi, Huang, Uribe-Romo, Chae,
  O’Keeffe, and Yaghi]{10.1073/pnas.0602439103}
Kyo~Sung Park, Zheng Ni, Adrien~P. C\^oté, Jae~Yong Choi, Rudan Huang,
  Fernando~J. Uribe-Romo, Hee~K. Chae, Michael O’Keeffe, and Omar~M. Yaghi.
\newblock Exceptional chemical and thermal stability of zeolitic imidazolate
  frameworks.
\newblock \emph{Proceedings of the National Academy of Sciences}, 103\penalty0
  (27):\penalty0 10186–10191, July 2006.
\newblock ISSN 1091-6490.
\newblock \doi{10.1073/pnas.0602439103}.
\newblock URL \url{http://dx.doi.org/10.1073/pnas.0602439103}.

\bibitem[Banerjee et~al.(2009)Banerjee, Furukawa, Britt, Knobler, O’Keeffe,
  and Yaghi]{10.1021/ja809459e}
Rahul Banerjee, Hiroyasu Furukawa, David Britt, Carolyn Knobler, Michael
  O’Keeffe, and Omar~M. Yaghi.
\newblock Control of pore size and functionality in isoreticular zeolitic
  imidazolate frameworks and their carbon dioxide selective capture properties.
\newblock \emph{Journal of the American Chemical Society}, 131\penalty0
  (11):\penalty0 3875–3877, March 2009.
\newblock ISSN 1520-5126.
\newblock \doi{10.1021/ja809459e}.
\newblock URL \url{http://dx.doi.org/10.1021/ja809459e}.

\bibitem[Hjorth~Larsen et~al.(2017)Hjorth~Larsen, Jørgen~Mortensen, Blomqvist,
  Castelli, Christensen, Dułak, Friis, Groves, Hammer, Hargus, Hermes,
  Jennings, Bjerre~Jensen, Kermode, Kitchin, Leonhard~Kolsbjerg, Kubal,
  Kaasbjerg, Lysgaard, Bergmann~Maronsson, Maxson, Olsen, Pastewka, Peterson,
  Rostgaard, Schiøtz, Sch\"{u}tt, Strange, Thygesen, Vegge, Vilhelmsen,
  Walter, Zeng, and Jacobsen]{10.1088/1361-648X/aa680e}
Ask Hjorth~Larsen, Jens Jørgen~Mortensen, Jakob Blomqvist, Ivano~E Castelli,
  Rune Christensen, Marcin Dułak, Jesper Friis, Michael~N Groves, Bjørk
  Hammer, Cory Hargus, Eric~D Hermes, Paul~C Jennings, Peter Bjerre~Jensen,
  James Kermode, John~R Kitchin, Esben Leonhard~Kolsbjerg, Joseph Kubal,
  Kristen Kaasbjerg, Steen Lysgaard, Jón Bergmann~Maronsson, Tristan Maxson,
  Thomas Olsen, Lars Pastewka, Andrew Peterson, Carsten Rostgaard, Jakob
  Schiøtz, Ole Sch\"{u}tt, Mikkel Strange, Kristian~S Thygesen, Tejs Vegge,
  Lasse Vilhelmsen, Michael Walter, Zhenhua Zeng, and Karsten~W Jacobsen.
\newblock The atomic simulation environment—a python library for working with
  atoms.
\newblock \emph{Journal of Physics: Condensed Matter}, 29\penalty0
  (27):\penalty0 273002, June 2017.
\newblock ISSN 1361-648X.
\newblock \doi{10.1088/1361-648x/aa680e}.
\newblock URL \url{http://dx.doi.org/10.1088/1361-648X/aa680e}.

\bibitem[Widom(1963)]{10.1063/1.1734110}
B.~Widom.
\newblock Some topics in the theory of fluids.
\newblock \emph{The Journal of Chemical Physics}, 39\penalty0 (11):\penalty0
  2808–2812, December 1963.
\newblock ISSN 1089-7690.
\newblock \doi{10.1063/1.1734110}.
\newblock URL \url{http://dx.doi.org/10.1063/1.1734110}.

\bibitem[Frenkel and Smit(2002)]{FrenkelSmit2002}
Daan Frenkel and Berend Smit.
\newblock \emph{Understanding Molecular Simulation: From Algorithms to
  Applications}.
\newblock Academic Press, San Diego, CA, USA, 2 edition, 2002.
\newblock ISBN 978-0122673511.

\bibitem[{facebookresearch}(2025)]{fairchem}
{facebookresearch}.
\newblock Fairchem: Fair chemistry's library of machine learning methods for
  chemistry.
\newblock \url{https://github.com/facebookresearch/fairchem}, 2025.

\bibitem[Batatia et~al.(2022{\natexlab{a}})Batatia, Kovacs, Simm, Ortner, and
  Csanyi]{Batatia2022mace}
Ilyes Batatia, David~Peter Kovacs, Gregor N.~C. Simm, Christoph Ortner, and
  Gabor Csanyi.
\newblock {MACE}: Higher order equivariant message passing neural networks for
  fast and accurate force fields.
\newblock In Alice~H. Oh, Alekh Agarwal, Danielle Belgrave, and Kyunghyun Cho,
  editors, \emph{Advances in Neural Information Processing Systems},
  2022{\natexlab{a}}.
\newblock URL \url{https://openreview.net/forum?id=YPpSngE-ZU}.

\bibitem[Batatia et~al.(2022{\natexlab{b}})Batatia, Batzner, Kov{\'a}cs,
  Musaelian, Simm, Drautz, Ortner, Kozinsky, and Cs{\'a}nyi]{Batatia2022Design}
Ilyes Batatia, Simon Batzner, D{\'a}vid~P{\'e}ter Kov{\'a}cs, Albert Musaelian,
  Gregor N.~C. Simm, Ralf Drautz, Christoph Ortner, Boris Kozinsky, and
  G{\'a}bor Cs{\'a}nyi.
\newblock The design space of e(3)-equivariant atom-centered interatomic
  potentials, 2022{\natexlab{b}}.

\bibitem[{Hugging Face, Inc.}(2026)]{huggingface}
{Hugging Face, Inc.}
\newblock {Hugging Face}.
\newblock \url{https://huggingface.co}, 2026.
\newblock Accessed: 2026-02-12.

\bibitem[Willems et~al.(2012)Willems, Rycroft, Kazi, Meza, and
  Haranczyk]{zeo_paper}
Thomas~F. Willems, Chris~H. Rycroft, Michaeel Kazi, Juan~C. Meza, and Maciej
  Haranczyk.
\newblock Algorithms and tools for high-throughput geometry-based analysis of
  crystalline porous materials.
\newblock \emph{Microporous and Mesoporous Materials}, 149\penalty0
  (1):\penalty0 134–141, February 2012.
\newblock ISSN 1387-1811.
\newblock \doi{10.1016/j.micromeso.2011.08.020}.
\newblock URL \url{http://dx.doi.org/10.1016/j.micromeso.2011.08.020}.

\bibitem[Sriram et~al.(2025)Sriram, Brabson, Yu, Choi, Abdelmaqsoud, Moubarak,
  de~Haan, L\"{o}we, Brehmer, Kitchin, Welling, Zitnick, Ulissi, Medford, and
  Sholl]{10.48550/arXiv.2508.03162}
Anuroop Sriram, Logan~M. Brabson, Xiaohan Yu, Sihoon Choi, Kareem Abdelmaqsoud,
  Elias Moubarak, Pim de~Haan, Sindy L\"{o}we, Johann Brehmer, John~R. Kitchin,
  Max Welling, C.~Lawrence Zitnick, Zachary Ulissi, Andrew~J. Medford, and
  David~S. Sholl.
\newblock The open dac 2025 dataset for sorbent discovery in direct air
  capture, 2025.
\newblock URL \url{https://arxiv.org/abs/2508.03162}.

\bibitem[Barroso-Luque et~al.()Barroso-Luque, Shuaibi, Fu, Wood, Dzamba, Gao,
  Rizvi, Zitnick, and Ulissi]{10.48550/arXiv.2410.12771}
Luis Barroso-Luque, Muhammed Shuaibi, Xiang Fu, Brandon~M. Wood, Misko Dzamba,
  Meng Gao, Ammar Rizvi, C.~Lawrence Zitnick, and Zachary~W. Ulissi.
\newblock Open materials 2024 (omat24) inorganic materials dataset and models.
\newblock \doi{10.48550/arXiv.2410.12771}.
\newblock URL \url{http://arxiv.org/abs/2410.12771}.

\bibitem[mpt(2023)]{mptrj_2023}
{Materials Project Trajectory (MPtrj) dataset}.
\newblock
  \url{https://figshare.com/articles/dataset/Materials_Project_Trjectory_MPtrj_Dataset/23713842},
  2023.
\newblock Figshare dataset (MPtrj) — accessed 2025-10-30.

\bibitem[sal(2024)]{salex_2024}
{sAlex: a Matbench-Discovery compliant subsample of the Alexandria dataset}.
\newblock \url{https://matbench-discovery.materialsproject.org/data/salex},
  2024.
\newblock sAlex dataset used in MACE second-generation models. Accessed
  2025-10-30.

\end{thebibliography}

\end{document}